\documentclass[sn-basic]{sn-jnl}

\usepackage{natbib}
\usepackage{units}

\jyear{2023}%

\theoremstyle{thmstyleone}%

\theoremstyle{thmstyletwo}%

\theoremstyle{thmstylethree}%

\begin{document}

\title[Stellar Activity Cycles]{Stellar Activity Cycles}


\author*[1]{\fnm{Sandra V.} \sur{Jeffers}}\email{jeffers@mps.mpg.de}

\author[2]{\fnm{Ren\'e} \sur{Kiefer}}\email{kiefer@leibniz-kis.de}

\author[3]{\fnm{Travis S.} \sur{Metcalfe}}\email{travis@wdrc.org}

\affil*[1]{\orgname{Max Planck Institut für Sonnensystemforschung}, \orgaddress{\street{Justus-von-Liebig-Weg 3}, \city{G\"ottingen}, \postcode{37077},  \country{Germany}}}

\affil[2]{\orgname{Leibniz Institute for Solar Physics (KIS)}, \orgaddress{\street{Sch\"oneckstra\ss e 6}, \city{Freiburg}, \postcode{79104}, \country{Germany}}}

\affil[3]{\orgname{White Dwarf Research Corporation}, \orgaddress{\street{9020 Brumm Trail}, \city{Golden}, \state{Colorado} \postcode{80403}, \country{USA}}}

\abstract{The magnetic field of the Sun is generated by internal dynamo process with a cyclic period of 11 years or a 22 year magnetic cycle.  The signatures of the Sun's magnetic cycle are observed in the different layers of its atmosphere and in its internal layers.  In this review, we use the same diagnostics to understand the magnetic cycles of other stars with the same internal structure as the Sun. We review what is currently known about mapping the surface magnetic fields, chromospheric and coronal indicators, cycles in photometry and asteroseismology.  We conclude our review with an outlook for the future.}


\maketitle

\section{Introduction}\label{sec1}

Magnetic fields on the Sun are well characterised with observations at both high spatial and temporal resolutions. The signatures of the solar dynamo are observable in the different layers of the Sun's atmosphere using multi-wavelength observations and through seismology that allows us to directly probe beneath the Sun's surface. For a comprehensive review of the solar dynamo we refer to the recent review of \cite{Charbonneau2020LRSP...17....4C}.  For stars with masses ranging from slightly higher than the Sun down to about one-third of the Sun's mass, they are known to have a comparable internal structure as the Sun, in the form of an internal radiative zone and an convective envelope. Since the presence of an outer convective envelope is a key ingredient in the generation of the Solar dynamo, we also expect to see comparable signatures of magnetic activity on other solar-type stars. While it is unfortunately not possible to observe other stars with the same spatial resolution or temporal cadence as the Sun, they do allow us to understand how key components of the solar dynamo, such as stellar mass and rotation rate, impact the internal dynamo processes. 

A key component of the solar dynamo is that the dynamo mechanism produces a balance between the amounts of magnetic flux generated and lost over its 11-year activity cycle.   At the beginning of the cycle, magnetic flux emerges at mid-latitudes on the Sun's surface. Over the course of the cycle, the latitude of flux emergence decreases, reflecting the changing nature of the Sun's internal magnetic field, and finally reaches the Sun's equator. This winged pattern of flux emergence is commonly depicted in a solar butterfly diagram \citep{Maunder1904MNRAS..64..747M}.   The importance of the surface magnetic fields in the dynamo process has been recently reviewed by \cite{Cameron2023arXiv230502253C}.  The first detection of a magnetic field in sunspots was reported by \cite{1908ApJ....28..315H}. The evolution of the Sun's large-scale magnetic field over its cycle has been monitored by regular polarimetric observations since the 1970s \citep{Cameron2018A&A...609A..56C}. For each 11-year cycle, the magnetic field changes polarity, leading to a 22-year magnetic cycle \citep{Hathaway2015LRSP...12....4H}. Higher up in the Sun's atmosphere, chromospheric diagnostics are commonly used such as the $S$ index. The $S$ index is a measure of the emission in the cores of the Ca~{\sc ii} H and K lines, at \unit[396.6]{nm} and \unit[393.4]{nm} respectively, relative to nearby spectral continuum regions \citep{Wilson1968ApJ...153..221W}. Another key feature of the solar dynamo is that the Sun's $S$ index is co-incident with the evolution of the geometry of the solar large-scale magnetic field, where a complex geometry occurs at $S$ index activity maximum and a more simple dipole at $S$ index activity minimum. Similarly, the Sun's coronal emission variations, as observed in X-rays, are also co-incident with the evolution of the Sun's large-scale magnetic field \citep{Ayres2020}.   While \cite{Radick2018ApJ...855...75R} reported that the photometric variations of the Sun are co-incident with variations in its Ca~{\sc ii} emission and the total solar irradiance.  

Probing the internal structure of the Sun using the technique of helioseismology allows us to understand how the internal layers of the Sun are impacted by its magnetic cycle. The primary diagnostic is via globally resonant acoustic waves, or p-modes. The specific way in which the p-modes' parameters vary as a consequence of the changing levels of magnetic activity and magnetic field strength, in particular the p-mode frequencies, encodes information about the perturbation causing these changes. In the Sun, the frequencies of p-modes are correlated with the level of magnetic activity, whereas p-mode amplitudes are anti-correlated with the level of magnetic activity. While there are many more indicators to characterise the magnetic cycle of the Sun, these are the main seismic diagnostics that can be searched for in the photometric time series of other stars and can yield insights into their magnetic cycles. 

One of the challenges in observing and characterising stellar cycles is the long timespans that are needed to acquire definitive results. The longest data sets are those of the $S$ index for the Sun and for other stars. The longest continuous sets of $S$ index chromospheric activity measurements were obtained over 35~years at the Mount Wilson observatory\footnote{For publicly available data, see \url{https://dataverse.harvard.edu/dataverse/mwo\_hk\_project}} between 1968 and 2003 \citep{Olah2016}, although most of the published datasets end in 1992 \citep{Baliunas1995}. Additional long-term observations are available from Lowell Observatory \citep[1992--2020;][]{Hall2007}, Keck Observatory \citep[1996--present;][]{Baum2022}, the TIGRE telescope \citep[2013--present;][]{Gonzalez2022}, ESPaDOnS (Echelle SpectroPolarimetric Device for the Observation of STARS) \citep[2006--present;][]{Brown2022MNRAS.514.4300B} and NARVAL \citep[2007--present;][]{Brown2022MNRAS.514.4300B} in the northern hemisphere, and from ESO 
\citep[2003--present;][]{Lovis2011} and SMARTS \citep[2007--2013;][]{Metcalfe2009} in the southern hemisphere. 

In terms of understanding stellar cycles, there has recently been significant progress in understanding how the geometry of a star's large-scale magnetic field varies over its magnetic cycle. This is due to the development of instrumentation to reconstruct, and dedicated telescopes to monitor, the large-scale magnetic fields of many stars. The main instruments used to observe the magnetic fields of stars are: (1) ESPaDOnS at the \unit[3.6]{m} Canada France Hawaii telescope \citep{Donati2006ASPC..358..362D}, its twin (2) NARVAL at the \unit[2.2]{m} Telescope Bernard Lyot \citep[TBL;][]{Auriere2003EAS.....9..105A}, and (3) HARPSpol \citep{Snik2008SPIE.7014E..0OS,Piskunov2011Msngr.143....7P}. The advantage of ESPaDOnS and NARVAL is that they were specifically designed with the purpose of monitoring the magnetic fields of stars. For example, their long wavelength ranges cover wavelengths from the far-UV to the nIR and are capable of observing many thousands of lines that can be combined to increase the information content using techniques such as least-squares deconvolution \citep[LSD;][]{Donati1997MNRAS.291..658D, Kochukhov2010A&A...524A...5K}. Additionally, NARVAL is the only instrument at the TBL allowing detailed monitoring of the large-scale magnetic field of stars on short to long timescales. Even though these instrumental developments were commissioned more than 15 years ago, it is only now that we have a sufficiently long time span of data to understand the intrinsic variability of these stars on timescales of the order of the solar cycle.

Furthermore, the field of asteroseismology has rapidly advanced over the last 15 years with the advent of high-precision photometric missions such as CoRoT \citep{Baglin2006, Auvergne2009}, \textit{Kepler} \citep{Borucki2010, Koch2010} and currently TESS \citep{Ricker2014}. Data from these space missions secured with a high cadence has allowed us to put the Sun in the context of other stars and to monitor the impact of stellar magnetic fields on the stars' internal layers. 

In this review article, we focus on the observational diagnostics of stellar magnetic cycles of F, G and K dwarfs. We first address photospheric signatures of stellar cycles in Section~\ref{sec2}, and then observations of cycles in the stellar chromosphere and corona in Section~\ref{sec3}. Finally, we present cycles in terms of the internal layers of stars via astroseismology in Section~\ref{sec4}. For completeness we also briefly summarise what is known about photometric cycles in Section~\ref{sec5}. We then discuss the future prospects in Section~\ref{sec6}.  

\section{Photospheric diagnostics}\label{sec2}

\begin{figure*}
\centering
\includegraphics[width=0.25\textwidth,height=5.2cm]{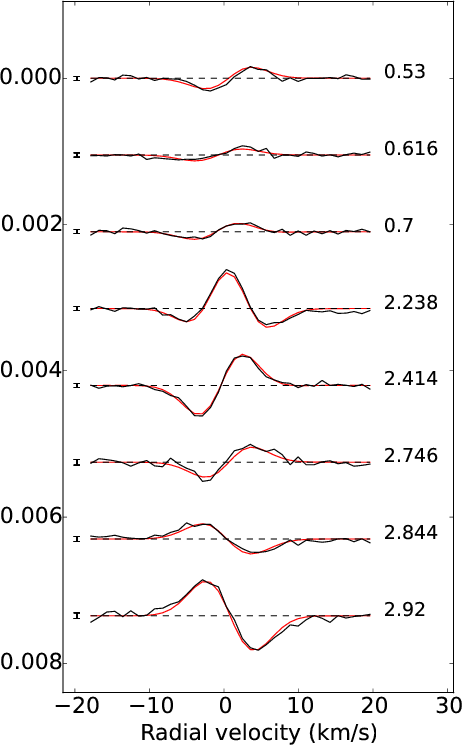}
\includegraphics[width=0.4\textwidth,height=5.25cm]{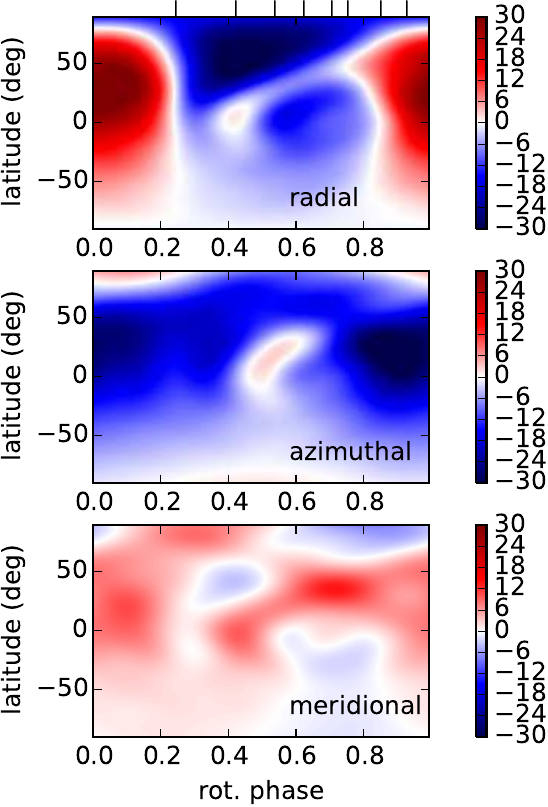}
\caption{Time-series of Stokes {\it{V}} profiles for $\epsilon$~Eri in October 2015 (\textit{left panel}) with the reconstructed large-scale magnetic field geometry shown in the three panels on the right. For the Stokes {\it{V}} profiles, the black lines represent the data and red dotted lines correspond to synthetic profiles of our magnetic model and where successive profiles are shifted vertically for display clarity. Rotational phases of observations are indicated in the right part of the Stokes V LSD profiles and as vertical ticks at the top of the ZDI map. Data is from the NARVAL spectropolarimeter at the 2.2m Telescope Bernard Lyot secured as part of the BCool collaboration.}
\label{fig:eps_eri_new}
\end{figure*}

To reconstruct the large-scale magnetic field of stars other than the Sun, a commonly used method is Zeeman-Doppler Imaging (ZDI). This technique can measure the net magnetic field from a non-homogeneous distribution of circularly polarised light as a function of stellar rotation phase. As previously mentioned, the information content of thousands of spectral lines is combined using the technique of LSD. From a time series of LSD Stokes V observations covering not more than a few stellar rotation periods, the ZDI technique inverts the observed circularly polarised profiles into the strength, polarisation and distribution on the stellar surface in terms of the poloidal, toroidal and meridional magnetic fields (via spherical harmonics expansion \citealp{Donati2006MNRAS.370..629D, Folsom2018MNRAS.474.4956F}). An example of the time series of Stokes V LSD profiles and the corresponding ZDI map is shown in Figure~\ref{fig:eps_eri_new} for the K dwarf $\epsilon$ Eri in October 2015.   In this review we specifically focus on the results from ZDI as these are many times more numerous than the tentative results from Doppler imaging alone \cite[for example][for the long-term brightness monitoring of AB Dor and II Peg]{Jeffers2007MNRAS.375..567J,Hackman2012A&A...538A.126H}.

\begin{figure*}
\centering
\includegraphics[angle=0,width=1.0\linewidth]{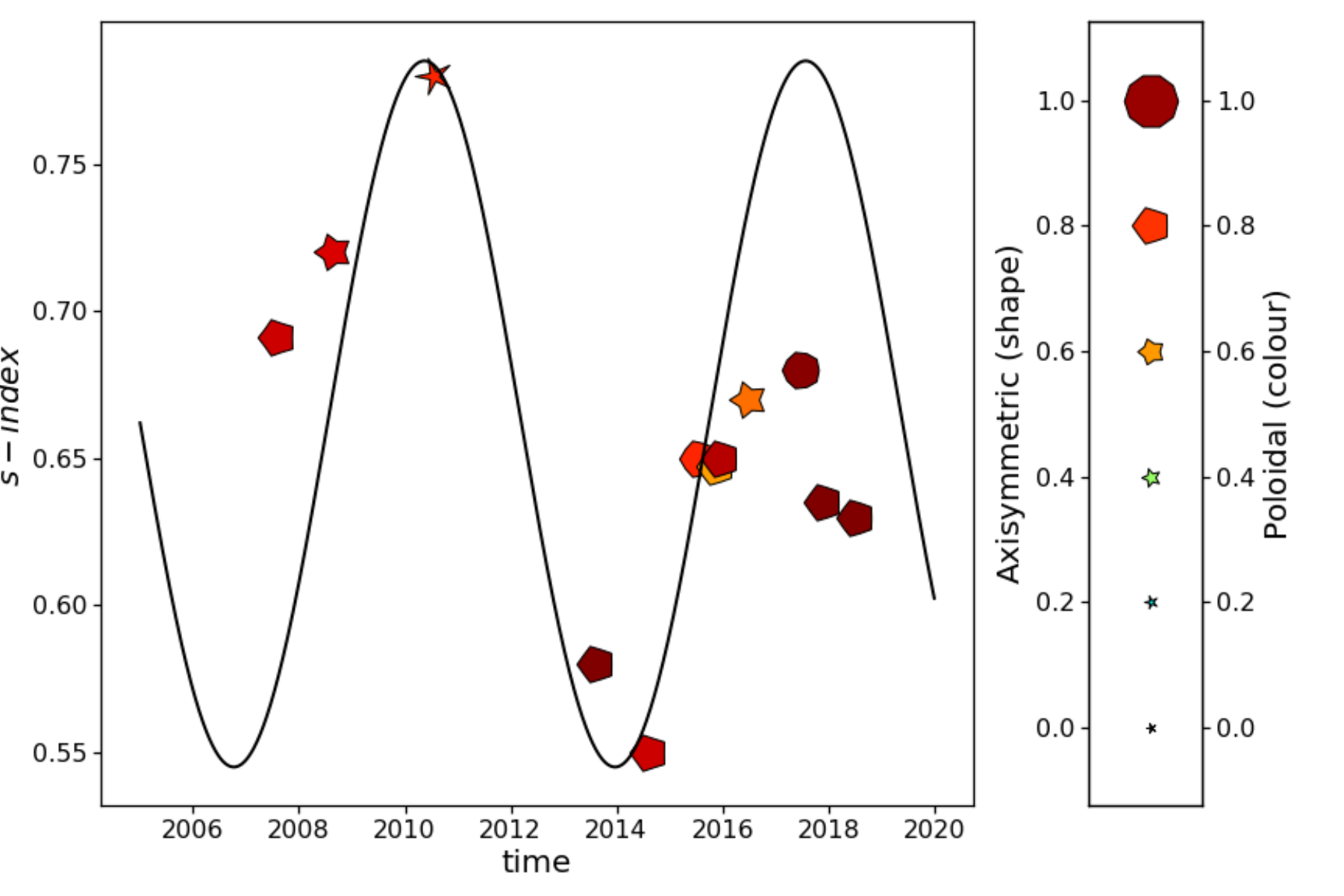}
\caption{Magnetic and $S$ index cycle of 61~Cyg~A as a function of time. The solid black line is a sinusoid with the period 61~Cyg~A's $S$ index cycle of \unit[7.3]{yr}. Each of the coloured data points summarises the magnetic field geometry of one magnetic map. Data points are coloured depending on whether the large-scale field is poloidal (dark red) or toroidal (blue). The shape of the points indicates the dipolar fraction of the poloidal field, with decagon shaped symbols having a high degree of axisymmetry. Data is from the NARVAL spectropolarimeter at the 2.2m Telescope Bernard Lyot secured as part of the BCool collaboration.}
\label{fig:61Cyg_confusogram}
\end{figure*}

\begin{figure}
\begin{center}
\includegraphics[width=0.8\linewidth]{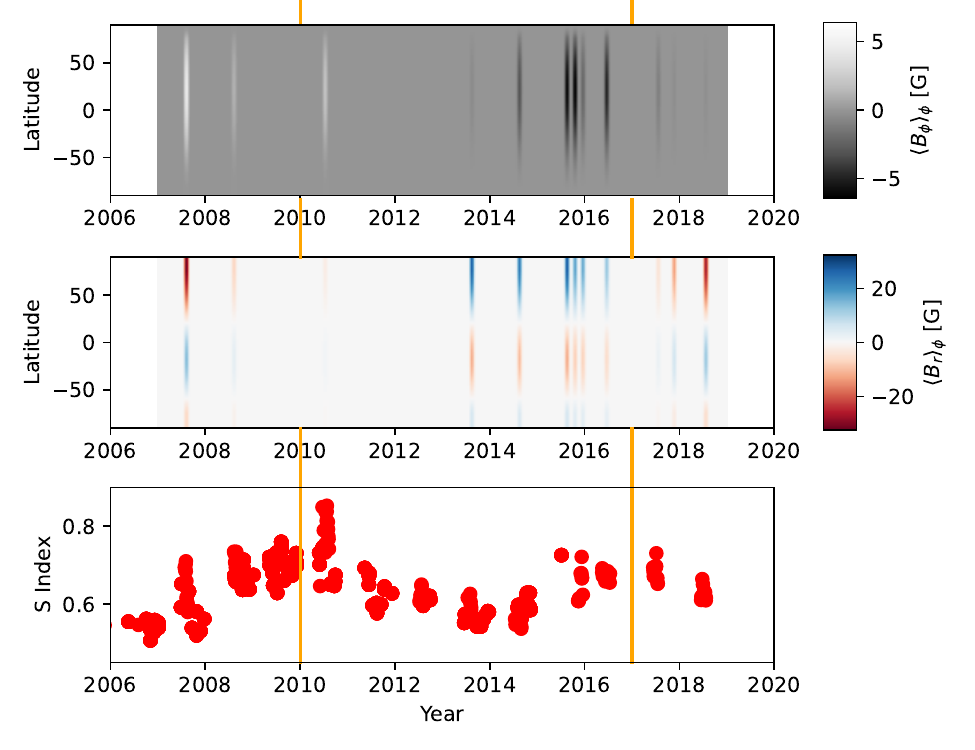}\\
\includegraphics[width=0.8\linewidth]{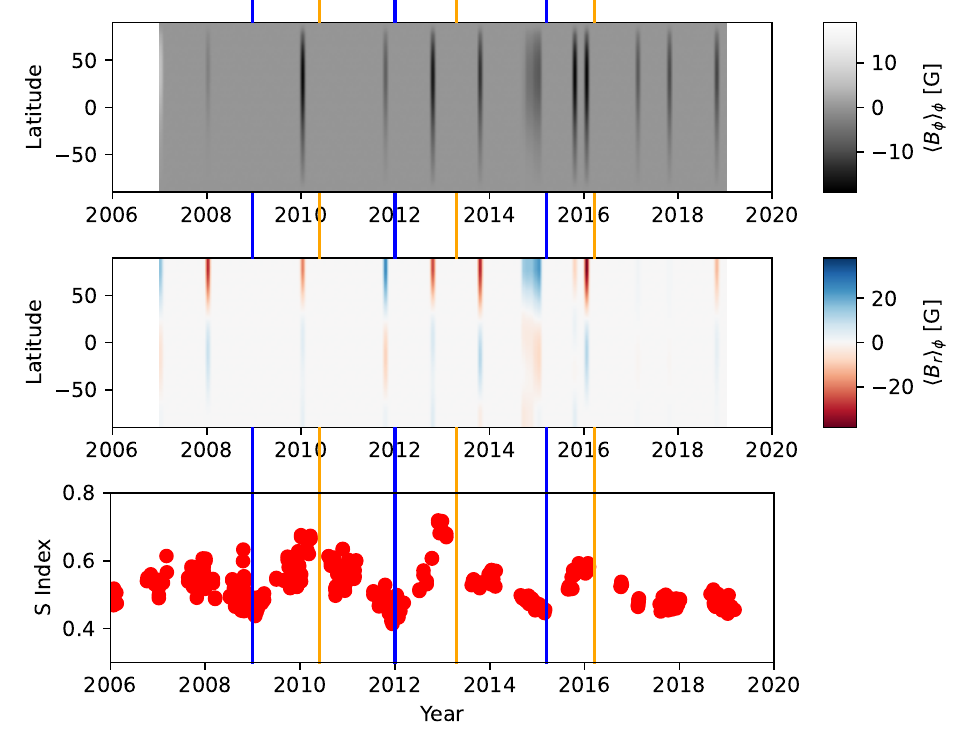}
\caption{Butterfly diagrams for 61~Cyg~A (upper three panels) and $\epsilon$~Eri (lower three panels). The vertical blue and orange lines indicate the $S$ index minima and maxima, respectively. Taken from \cite{Jeffers2022A&A...661A.152J}.
\label{fig:Butterfly-stars}
}
\end{center}
\end{figure}

A key diagnostic of the solar magnetic cycle is that the Sun's chromospheric cycle, or $S$ index cycle, is in phase with its magnetic cycle meaning that its large-scale magnetic field switches polarity at activity maximum. The first star reported to show a solar-like magnetic cycle is the K dwarf 61~Cyg~A \citep{BoroSaikia2016A&A...594A..29B, BoroSaikia2018A&A...620L..11B}. 61~Cyg~A is an old K5 dwarf, with a very slow rotation rate and a $S$ index cycle of \unit[7.3]{yr}. It shows polarity switches of its large-scale magnetic field in phase with its $S$ index activity maximum. Similar to the Sun, the large-scale field geometry is complex at activity maximum and dipolar at activity minimum. A summary of 61~Cyg~A's magnetic cycle is shown in Figure~\ref{fig:61Cyg_confusogram}. Another K dwarf with an extensive time span of magnetic maps is the young, rapidly rotating K dwarf, $\epsilon$~Eri which has been shown by \cite{Metcalfe2013} to have two $S$ index cycles of \unit[2.95]{yr} and \unit[12.7]{yr}. In contrast to 61~Cyg~A and the Sun, $\epsilon$~Eri's large-scale magnetic field geometry shows a high degree of complexity at the minimum of the shorter $\sim$\,3-year $S$ index cycle \citep{Jeffers2022A&A...661A.152J}. It also does not show a change in polarity with every $S$ index maximum \citep{Jeffers2014A&A...569A..79J, Petit2021A&A...648A..55P}. However, \cite{Jeffers2022A&A...661A.152J} recently showed that this could be explained if $\epsilon$~Eri's shorter $\sim$\,3-year cycle is a modulation of its longer $\sim$\,13-year cycle and that a polarity switch in its large-scale magnetic field should occur in phase with the longer $S$ index cycle (Figure~\ref{fig:Butterfly-stars}).  Indications of third cycle period has been recently been presented by \cite{Fuhrmeister2023A&A...672A.149F}, though long-term observations of $\epsilon$~Eri's S-index are needed to confirm this additional cycle.  Other young K dwarfs that have been monitored using multi-epoch ZDI observations include the very young stars AB Dor and LQ Hya \citep{Donati2003MNRAS.345.1145D}, though no clear cyclic behaviour was identified.  More recently, \cite{Lehtinen2022A&A...660A.141L} reported a polarity reversal in LQ Hya, that is coincident with a possible S-index activity minimum.

For G dwarfs that are approximately within \unit[10]{\%} of the Sun's mass, we have a long timespan of magnetic maps for HD~171488 \citep{Marsden2006MNRAS.370..468M, Jeffers2008MNRAS.390..635J, Jeffers2011MNRAS.411.1301J, Willamo2022arXiv220313398W}, HN~Peg \citep{BoroSaikia2015A&A...573A..17B}, EK~Dra \citep{Waite2017MNRAS.465.2076W}, HD~190771 \citep{Petit2009A&A...508L...9P, Morgenthaler2011AN....332..866M} and $\kappa$~Cet \citep{BoroSaikia2022A&A...658A..16B}. However, we are yet to observe a solar-like magnetic cycle where the large-scale magnetic field switches polarity at $S$ index maximum on another early-G dwarf. This is .epsly because the G dwarfs that have been investigated with multi-epoch maps all are much younger than the Sun. For the three young stars EK Dra, HD~171488, and HN~Peg the large-scale magnetic field evolves rapidly with little evidence of polarity switches in either the poloidal or toroidal fields. In particular, for HN~Peg, the toroidal field appears and disappears again without any correlation with other activity indicators such as the $S$ index.  Polarity reversals have been observed in the G2 dwarf HD~190771 \citep{Petit2009A&A...508L...9P, Morgenthaler2011AN....332..866M}, however, they appear first in the azimuthal field \citep{Petit2009A&A...508L...9P} but are not observed in subsequent observations \citep{Morgenthaler2011AN....332..866M}. Somewhat surprisingly, \cite{Morgenthaler2011AN....332..866M} report a polarity reversal in subsequent epochs of HD~190771's radial field. Recently, \cite{BoroSaikia2022A&A...658A..16B} reported a potential $\sim$\,10-year magnetic cycle for the G5\,V star $\kappa$~Cet, a moderately active star with a rotation period of \unit[9.2]{d}. $\kappa$~Cet is also slightly older than EK~Dra, HD~171488, and HN~Peg with an age of $\sim$\,\unit[750]{Myrs}. Similar to the K2 dwarf $\epsilon$~Eri, $\kappa$~Cet shows evidence for having two chromospheric cycle periods of \unit[3.1]{yr} and \unit[6]{yr}, respectively. The longer period dominates for most of the data set analysed by \cite{BoroSaikia2022A&A...658A..16B}. Interestingly, the shorter period is present at the beginning and end of the dataset, when the longer period disappears. Additional spectropolarimetric observations that densely sample $\kappa$~Cet's $S$ index cycle periods will help to resolve the intriguing case of the cyclic evolution of $\kappa$~Cet's magnetic field.

The first star to show successive polarity switches in its large-scale magnetic field geometry is the planet-hosting late-F dwarf $\tau$~Boo \citep{Catala2007MNRAS.374L..42C, Donati2008MNRAS.385.1179D, Fares2009MNRAS.398.1383F, Fares2013MNRAS.435.1451F}. With a rotation period of only \unit[3.3]{d}, $\tau$~Boo is more massive and has a much faster rotation than 61~Cyg~A. The first results indicated a polarity switch occurring between June 2006 and June 2007, and then again between June 2007 and July 2008 from which \cite{Fares2013MNRAS.435.1451F} concluded that the magnetic cycle was \unit[2]{yr} long. Subsequent work by \cite{Mengel2016MNRAS.459.4325M} presented additional epochs of $\tau$~Boo's large-scale magnetic field and reported that $\tau$~Boo has an $S$ index cycle length of \unit[120]{d}. More recently, the densely sampled observations by \cite{Jeffers2018MNRAS.479.5266J} over $\tau$~Boo's $S$ index cycle showed that $\tau$~Boo's magnetic cycle is indeed co-incident with its $S$ index cycle and with a polarity reversal at $S$ index activity maximum. This makes $\tau$~Boo one of the shortest magnetic cycles with a length of \unit[240]{d}. $\tau$~Boo's giant $\sim$\,6 Jupiter-mass planet, which orbits at a distance of $\sim\,\unit[0.5]{AU}$, has been considered to play a role in its internal dynamo processes via tidal locking, similar to the increased activity of stars in binary systems (see \citealp{Fares2013MNRAS.435.1451F}). Recent work by \cite{Brown2021MNRAS.501.3981B} showed that similar levels of activity and polarity reversals also exist on HD~75332 a star with a similar mass and rotation rate as $\tau$~Boo. Using 12 epochs of magnetic maps, \cite{Brown2021MNRAS.501.3981B} showed that HD~75332 has a rapid 1.06 yr cycle and that a polarity reversal at activity maximum is consistent with polarity switches at activity maximum but a repeated polarity switch is required to confirm the cyclic nature of HD~75332's large-scale magnetic field. Another late-F star that shows evidence for a potential 3-year cycle is HD~78366 \citep{Morgenthaler2011AN....332..866M}, where the radial field shows polarity reversals with a possible 3-year cycle.  In contrast, the recently work of \cite{Marsden2023MNRAS.522..792M} showed that the large-scale magnetic field of the old F7 dwarf Chi Dra is stable over a 5 year time span and does not show indications of cyclic behaviour.     The results from these stars show the that rapid cyclic nature of the large-scale magnetic field of late F-stars could be an intrinsic feature of young to middle aged stars with a shallow convective zone, and that more stable patterns emerge as these stars evolve off the .eps-sequence.  

While the observational data is not yet conclusive on several targets, the dense phase coverage of 61~Cyg~A and $\epsilon$~Eri allow a more detailed comparison with the workings of the solar dynamo processes. In the case of the Sun, it is well established that the axisymmetric component of the 
toroidal field $\langle B_\phi\rangle$ \citep{Cameron2018A&A...609A..56C} is a proxy for flux emergence of the global dynamo and follows the Sun's $S$ index. Recent work by \cite{Jeffers2022A&A...661A.152J} shows that this relation still holds at the resolution of the magnetic maps reconstructed with ZDI. Applying this to 61~Cyg~A shows that the flux emergence also follows its $S$ index, while for $\epsilon$~Eri it shows two cycles and the potential onset of an extended inactive period. The work of \cite{Jeffers2022A&A...661A.152J} concludes that surface magnetic fields play a crucial role in the dynamos of 61~Cyg~A, $\epsilon$~Eri, and the Sun.  For further discussion on the nature of the stellar dynamo from a modelling perspective we refer to \cite{Brun2022ApJ...926...21B,Petri2023arXiv230516790K} and references therein.  

\section{Chromospheric and Coronal Diagnostics}\label{sec3}

Most of the available data on stellar activity cycles come from observations of the 
Ca~{\sc ii} H (\unit[396.6]{nm}) and K (\unit[393.4]{nm}) spectral lines (hereafter Ca~HK). Emission in 
the cores of these lines is a well-established proxy for magnetic heating in the 
chromosphere \citep{Leighton1959}. Time series observations with an appropriate cadence 
can probe both the long-term variations due to magnetic activity cycles 
\citep{Wilson1978}, as well as shorter-term modulation due to stellar rotation 
\citep{Baliunas1983}. Considering the 11-year sunspot cycle, decades of observations are 
typically required to measure activity cycles in other stars.  A list of stars with currently known chromospheric cycles was recently compiled by \cite{Mittag2023A&A...674A.116M} (see their Table 1).

\begin{figure}[t]
\centering\includegraphics[width=0.9\textwidth]{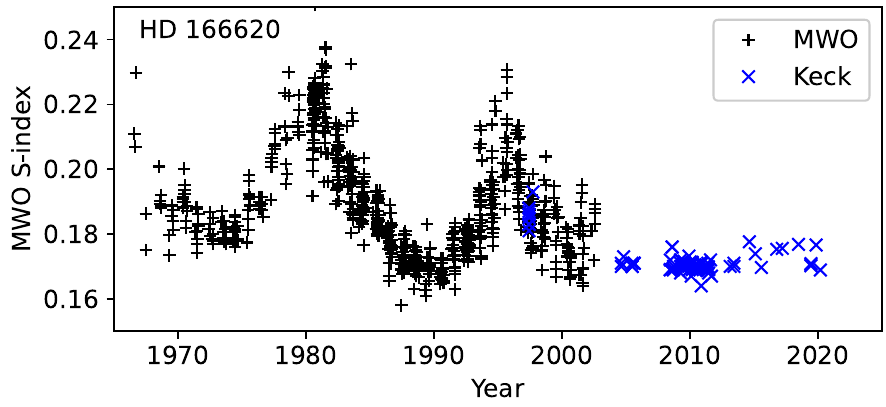}
\caption{Stellar activity measurements for the old K-type star HD~166620 spanning more 
than 50 years, including observations from Mount Wilson (black points) and Keck (blue 
points). This is the first unambiguous example of a Sun-like star entering a grand 
magnetic minimum \citep[data from][]{Olah2016, Baum2022}.\label{fig3.1}}
\end{figure}

As an illustration of what we can learn from multi-decadal time series measurements, 
the combined datasets from Mount Wilson and Keck for the old K-type star HD~166620 are 
shown in Fig.~\ref{fig3.1}. The Mount Wilson Observatory (MWO) $S$ index is one of the standard proxies for 
magnetic activity, measuring emission in the Ca~HK line cores relative to nearby
pseudo-continuum bands. The first few decades of observations from Mount Wilson 
\citep[black points in Fig.~\ref{fig3.1};][]{Olah2016} reveal a regular activity cycle with a period \mbox{$P_{\rm 
cyc}\!\sim\!\unit[15]{yr}$}. Higher-cadence observations beginning in the 1980s reveal 
rotational modulation with a mean period $P_{\rm rot}\!\sim\!\unit[42]{d}$, and significant 
differential rotation from seasonal variations (\unit[33.4--50.8]{d}), presumably as 
active regions migrate to different latitudes through the cycle \citep{Donahue1996}. 
In addition, the apparent correlation between cycle amplitude and the rise time for 
individual cycles has been used to examine whether the Waldmeier Effect in the Sun 
\citep{Waldmeier1935} is also observed for other Sun-like stars \citep{Garg2019, Willamo2020}.
Most notably, the continued observations from Keck \citep[blue points in Fig.~\ref{fig3.1};][]{Baum2022} 
reveal a smooth transition from cycling to constant activity, the first unambiguous 
example of a Sun-like star entering a grand magnetic minimum \citep{Luhn2022}. The 
high-cadence observations around 2010 coincide with the next expected maximum of the
cycle, which is clearly absent. This may support the idea that stellar cycles can 
become intermittent as stars evolve through the critical activity level where weakened 
magnetic braking appears to begin \citep{vanSaders2016, Metcalfe2022}. 

Some stellar cycles have also been observed at X-ray wavelengths. The magnetic processes
that heat the chromosphere also heat the corona, which has a much higher temperature
($\sim\unit[10^6]{K}$, emitting at X-ray wavelengths) and fills a larger volume around the
star. Variations around the mean X-ray luminosity are substantially larger
than variations around the mean Ca~HK emission, providing a higher contrast for the
detection of magnetic cycles. For example, the young solar-type star $\iota$~Hor shows
fractional variations that are 3--4 times larger in X-ray luminosity than in Ca~HK
emission \citep{SanzForcada2013}. The .eps challenge is that X-ray measurements must
be obtained from above the Earth's atmosphere, so the cadence and duration of the
observations are limited by competition for time on space telescopes and the longevity of \mbox{X-ray} missions (for instrument stability).
These challenges have effectively forced studies of coronal activity cycles to rely on
measurements from two long-lived X-ray missions, {\it XMM-Newton} and {\it Chandra}.

Most observations of coronal activity cycles have focused on stars with previously known
activity cycles from Ca~HK measurements. The earliest detections included the 7.3-year
cycle in 61~Cyg~A \citep{Hempelmann2006, Robrade2012} and the 8.2-year cycle in HD~81809
\citep{Favata2008, Orlando2017}, both of which appeared to be approximately in phase with
chromospheric variations. The discovery of substantially shorter chromospheric activity
cycles in the young solar-type stars $\iota$~Hor \citep{Metcalfe2010, IotaHorAlvarado2018MNRAS.473.4326A} and $\epsilon$~Eri
\citep{Metcalfe2013} provided new opportunities to study coronal activity cycles on
shorter timescales \citep{SanzForcada2013, Coffaro2020}, and revealed some fascinating
incongruities between chromospheric and coronal variations. Discoveries of previously
unknown activity cycles are currently limited to the southern hemisphere stars $\alpha$~Cen~A~\&~B
\citep{Robrade2012}, which were inaccessible to the Mount Wilson survey. Characterization
of the coronal activity cycle in the \unit[5.4]{Gyr} solar analog $\alpha$~Cen~A is of particular
interest as a constraint on the future of the 11-year solar cycle. The amplitude of its
19.2-year X-ray cycle is about one-third that of the Sun \citep{Ayres2020}, suggesting
that the solar cycle may be growing longer and weaker \citep{Metcalfe2017}.

\begin{figure}[t]
\centering\includegraphics[width=0.9\textwidth]{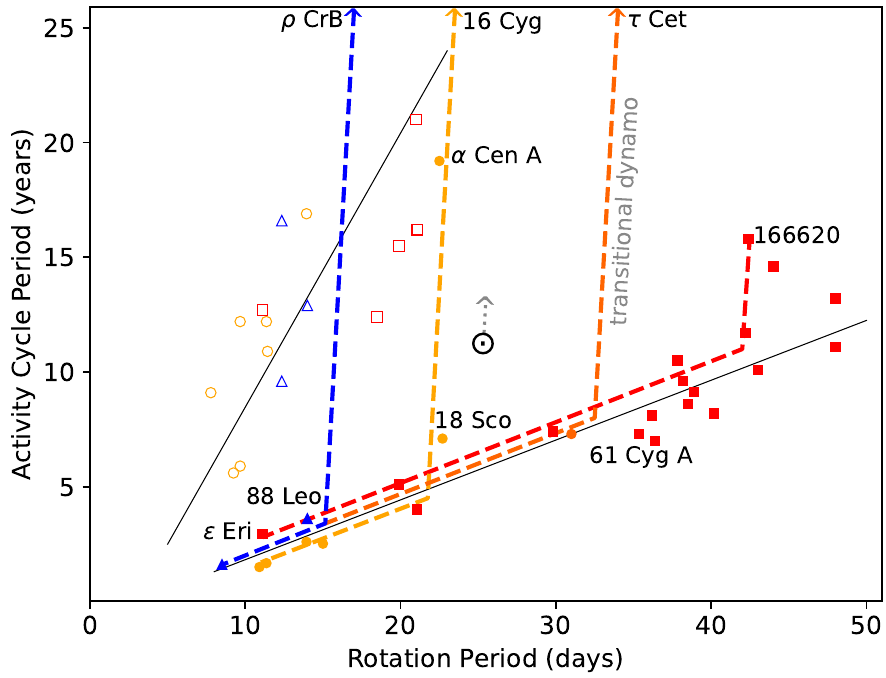}
\caption{Dependence of activity cycle period on rotation, showing two distinct sequences 
(solid lines). Points are coloured by effective temperature, indicating F-type (blue 
triangles), G-type (yellow and orange circles), and K-type stars (red squares). 
Schematic evolutionary tracks are shown as dashed lines, leading to 
stars with constant activity that appear to have shut down their global dynamos (arrows 
along top). Several notable stars are labelled.\label{fig3.2}}
\end{figure}

With high-quality measurements of stellar activity cycles and rotation periods, it is 
natural to ask whether there is any discernible relationship between these observables, 
as expected from dynamo theory. This question was examined empirically by Erika B\"ohm-Vitense in 2007 \citep{BohmVitense2007}. In the first figure of 
her thought-provoking paper, she simply plotted $P_{\rm cyc}$ against $P_{\rm rot}$ for 
stars in the Mount Wilson survey with the most reliable measurements \citep{Saar1999}. 
An updated version of this plot is shown in Fig.~\ref{fig3.2}, which reveals two 
distinct relationships between these two observables (solid lines). There is an upper 
sequence of long-period cycles \citep[open points; but see][]{BoroSaikia2018A&A...620L..11B}, and a lower sequence of short-period 
cycles (solid points), with the solar cycle falling curiously in between. Some of the 
stars exhibit cycles on both branches simultaneously, leading B\"ohm-Vitense to suggest 
that the two branches may represent two distinct dynamos operating in different regions 
of the star. Considering other properties of the stellar sample, she suggested that 
cycles on the long-period sequence may be driven in the near-surface shear layer, while 
the cycles on the short-period sequence may be driven at the base of the convection 
zone.

The first explanation for the peculiar position of the Sun in Fig.~\ref{fig3.2} came 
ten years later \citep{Metcalfe2017}. The updated version of the diagram coloured the points 
by spectral type, indicating hotter F-type (blue triangles), Sun-like G-type (yellow 
circles), and cooler K-type stars (red squares). The authors added several stars with 
measured rotation periods but no activity cycles (arrows along top) and included 
schematic evolutionary tracks (dashed lines) to indicate where rotation periods became 
nearly constant for different spectral types, apparently due to weakened magnetic 
braking \citep{vanSaders2016, Metcalfe2016}. According to this interpretation, activity 
cycles initially grow longer along each sequence as the stellar rotation period slows 
over time. However, when stars reach a critical Rossby number (the rotation period 
normalised by the convective turnover time), the rotation period re.epss nearly constant 
while the cycle gradually grows longer and weaker before disappearing entirely. This 
explains why hotter stars are confined to the left side of the diagram, while 
progressively cooler stars (with longer convective turnover times) continue to evolve 
further towards the right side. It also suggests that the Sun may be in a transitional 
evolutionary phase and that the solar cycle may represent a special case of dynamo theory.  These conclusions still hold even with the more recent results of \cite{BoroSaikia2018A&A...620L..11B} who confirm the lower sequence but question the presence of the upper sequence in Figure~\ref{fig3.2}.   Linking chromospheric cycles with the photospheric large-scale magnetic field (see Section 2), \cite{See2016MNRAS.462.4442S} reported that there are indications that stars on the upper sequence have highly variable toroidal fields, while stars on the lower branch have stable poloidal fields.  However, more stars with both chromospheric cycles and ZDI maps are needed to confirm this conclusion at short rotation periods.

An analysis contemporaneous to \cite{Metcalfe2017} by Axel Brandenburg and collaborators relied on a different 
representation of the measurements, more closely connected to dynamo theory 
\citep{Brandenburg2017}. Rather than plot $P_{\rm cyc}$ against $P_{\rm rot}$, the 
authors plotted $\log(P_{\rm rot}/P_{\rm cyc})$ (related to the strength of the $\alpha$ 
effect) against $\log R'_{\rm HK}$ \citep[related to the strength of the magnetic 
field;][]{Brandenburg1998}. The latter is the chromospheric emission from the MWO 
$S$ index, corrected for a small photospheric contribution and normalised by the 
bolometric luminosity of the star, allowing meaningful comparisons of stars with 
different spectral types. In this representation, a constant slope in a plot of $P_{\rm 
cyc}$ against $P_{\rm rot}$ becomes a horizontal line. However, the stellar data 
actually show a slope, indicating a weaker $\alpha$ effect as the magnetic field grows 
weaker. The solar cycle and HD~166620 both appear closer to the short-period sequence 
in this analysis, but the old solar analog $\alpha$~Cen~A \citep{Judge2017} and the 
K-type subgiant 94~Aqr~A \citep{Metcalfe2020} remain significant outliers.

\section{Seismology: insight into the internal structure}\label{sec4}
As the Sun and the stars pass through their activity cycles, the physical conditions in the regions in which the magnetic concentrations are located change over time. Solar and stellar oscillations propagating in these regions\footnote{p-, g-, and mixed modes; consult, e.g., the review by \citealp{Hekker2016} for more details about the different types of modes.} are sensitive to these changes. The specific way in which the modes parameters consequently vary, in particular the mode frequencies, contains valuable information about the perturbation causing these changes, i.e., the varying magnetic field. Therefore, helio- and asteroseismology enable us to probe the interior and atmospheric magnetic structure of the Sun and the stars.

\subsection{On the Sun}\label{sec4.1}
For the Sun, essentially all fundamental p-mode parameters are observed to vary over the solar activity cycle. The first parameter for which this was noticed, was the mode frequencies of low-degree modes \citep{Woodard1985}. This detection has been confirmed and expanded over the following decades \citep{Elsworth1990, Libbrecht1990, Jimenez-Reyes1998, 1999ApJ...524.1084H, Chaplin2001, Salabert2015a, Tripathy2015}. Now, the cyclic shift of p-mode frequencies, which is tightly in phase with the activity cycle for p-modes below the acoustic cut-off frequency, has been confirmed for a wide range of frequencies and harmonic degrees \citep[e.g.,][]{Broomhall2017}. In the context of stellar cycles, it is important to note that -- in contrast to p-mode frequencies, which are, as mentioned, correlated with the level of magnetic activity -- solar p-mode amplitudes are indeed anti-correlated with the level of magnetic activity \citep[e.g.,][]{Komm2000a, Jimenez2002a, JimenezReyes2003, JimenezReyes2004, Salabert2004, Burtseva2009, Broomhall2014, Broomhall2015, Kiefer2018a}. Also mode linewidths, which are related to mode damping \citep[e.g.,][]{Jefferies1991, Chaplin2000}, mode energies \citep[e.g.,][]{Komm2000, Kiefer2018a}, mode energy supply rates \citep{2021MNRAS.500.3095K}, and mode parameters of pseudomodes above the acoustic cut-off frequency \citep{2022MNRAS.512.5743K} vary through the solar cycle.

The sensitivity of p-modes to perturbations depends on their frequency as well as on their harmonic degree, increasing with both. This behaviour is largely due to the modes' inertia decreasing with both frequency and harmonic degree (see, e.g., \citealp{Christensen-Dalsgaard1991b, Komm2000, Chaplin2001} and for a more in-depth discussion and more references consult the review article by \citealp{Basu2016}). In contrast to the Sun, for stars, only the lowest harmonic degrees $l=0,1,2$ of p-modes can be measured, as the stellar photometric time series integrate the light of the full stellar disk. For these low harmonic degrees, the mode inertia does not differ very much between them. Any detected variation in the frequency shifts between modes of different harmonic degrees can be utilised to infer the latitudinal distribution of magnetic activity \citep{Moreno-Insertis2000, Chaplin2007, Thomas2021}. Further, mode frequency shifts, which increase with mode frequency, can be attributed to magnetic perturbations that are located very close to the surface, as higher frequency modes are concentrated to shallower layers \citep[e.g.,][]{Basu2012, Salabert2015a, Broomhall2017}.

\begin{figure}
 \centering
 \includegraphics[width=0.75\textwidth]{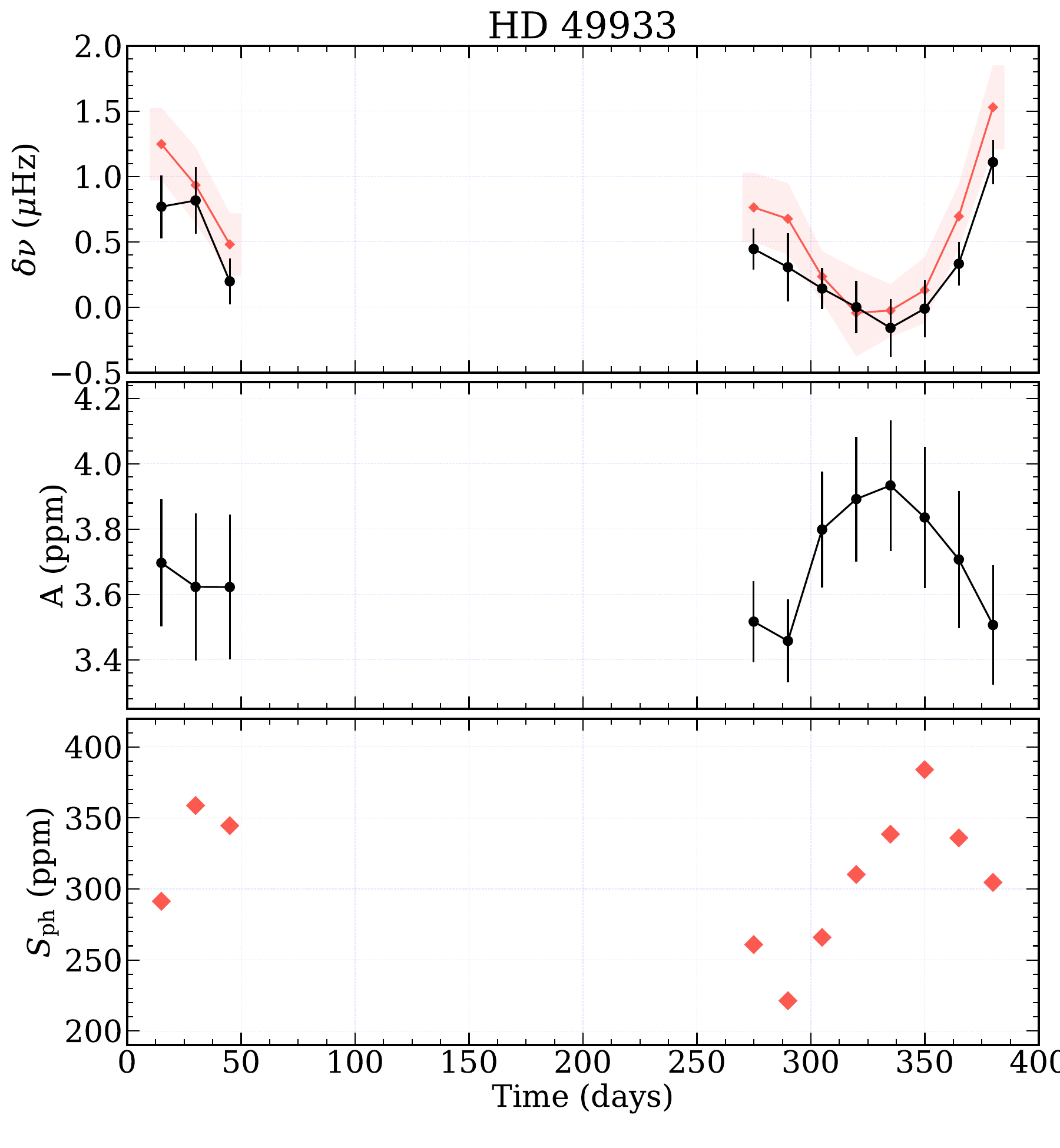}
 \caption{First detection of the asteroseismic fingerprint of stellar magnetic activity on the CoRoT target HD~49933 by \cite{Garcia2010}. \textit{Top panel:} p-mode frequency shifts $\delta\nu$ as measured via the cross-correlation (red diamonds) and the peak bagging approaches (black circles). \textit{Middle panel:} Mean p-mode amplitudes $A$ obtained from peak-bagging. \textit{Bottom panel:} Starspot proxy $S_{\mathrm{ph}}$ measured from the standard deviation of segments of the time series. This figure is a reproduction of Fig.~1 from \cite{Garcia2010} based on their original data.}
 \label{fig:garcia}
\end{figure}

\subsection{Asteroseismic detections of stellar magnetic activity (cycles)}\label{sec4.2}
It is the tight anti-correlation between the activity-related variations in p-mode frequencies and p-mode amplitudes that is a tell-tale seismic signature for varying levels of stellar magnetic activity in solar-like oscillators. This signature can be searched for in high-quality photometric time series that were delivered to us by the satellite missions CoRoT \citep{Baglin2006, Auvergne2009}, \textit{Kepler} \citep{Borucki2010, Koch2010}, and currently TESS \citep{Ricker2014}. If a stellar activity cycle is not fully covered by the data, the measured p-mode parameter variation generally presents a lower boundary on each star's cycle variability. In contrast, short activity cycles with periods of weeks or a few months may be missed by seismology: The length of the time series segments, that are needed to achieve the required frequency resolution to detect p-mode frequency variation on the order of a few tenths of \unit{\textmu Hz}, is typically around \unit[100]{d}.

\cite{Chaplin2007} and \cite{Karoff2009} investigated which types of stars ought to be observed and what characteristics the data must have if stellar activity cycles are to be detected and characterised seismically. \cite{Karoff2009} also include ground-based observations of chromospheric activity in their considerations. They found that, most importantly, the photometric time series as well as the ground-based observations need to be sufficiently long -- at least several consecutive months -- and the amplitude of the acoustic modes of the observed stars should be large enough so they protrude the noise \citep[also see][]{Chaplin2011, Chaplin2011a, Campante2016a, Ball2018, Schofield2019}. An in-depth review of the inferences asteroseismology can yield on stellar activity and activity cycles was also provided by \cite{Chaplin2014a}.

The first detection of activity-related p-mode parameter variations was achieved by \cite{Garcia2010} for the F-type dwarf star HD~49933 with CoRoT data. Their main results are reproduced in Fig.~\ref{fig:garcia}. The p-mode frequency shifts are depicted in the top panel as measured with two different methods (p-mode peak bagging in black circles, cross-correlation of the periodogram as red diamonds). The middle panel shows the p-mode amplitudes and the bottom panel shows a ``starspot proxy", which is the standard deviation of segments of the photometric time series. Indeed, the temporal changes in mode frequencies and mode amplitudes are clearly anti-correlated, pointing towards magnetic activity being the cause of these variations. As for the Sun, also HD~49933's frequency shifts increase with mode frequency, as \cite{Salabert2011} found. This indicates that the magnetic perturbation is located close to the star's surface.

Using \textit{Kepler} data of 24 solar-like stars with a length of at least \unit[960]{d}, \cite{Kiefer2017} found significant frequency shifts ($>1\sigma$) on 23 stars, showing that p-mode frequency variations are a very widespread phenomenon in solar-like oscillators. For six of these stars, the variation of p-mode amplitudes is also strongly anti-correlated (Spearman rank correlation coefficient $\rho<-0.5$) with the observed shifts of their frequencies. Shortly before, \cite{Salabert2016a} already found that the young solar analog KIC~10644253 exhibits activity-related p-mode frequency shifts. Based on spectroscopic observations with the HERMES spectrograph, they also demonstrated that this star is more active than the Sun.

\begin{figure}
 \centering
 \includegraphics[width=0.7\textwidth]{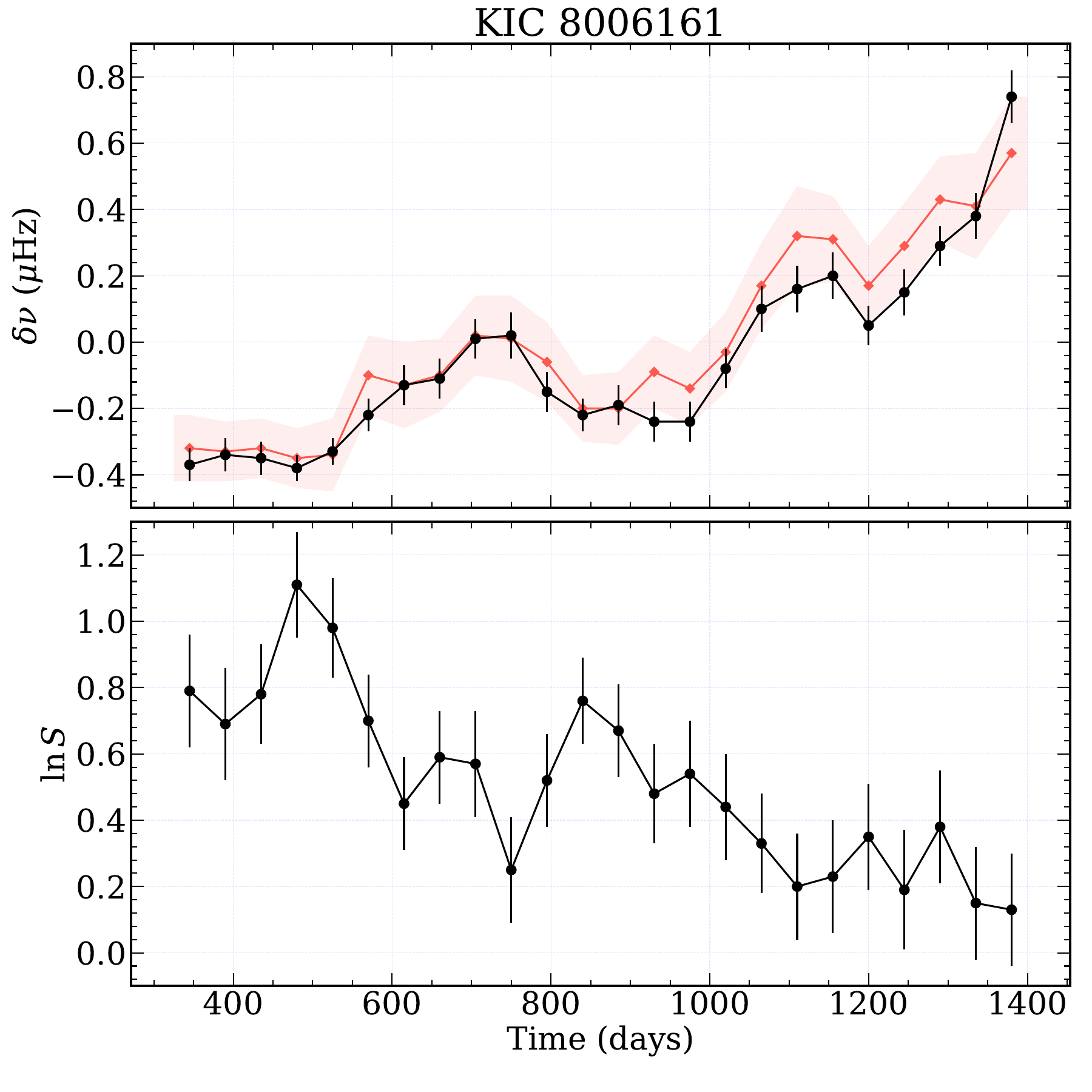}\\
 \includegraphics[width=0.7\textwidth]{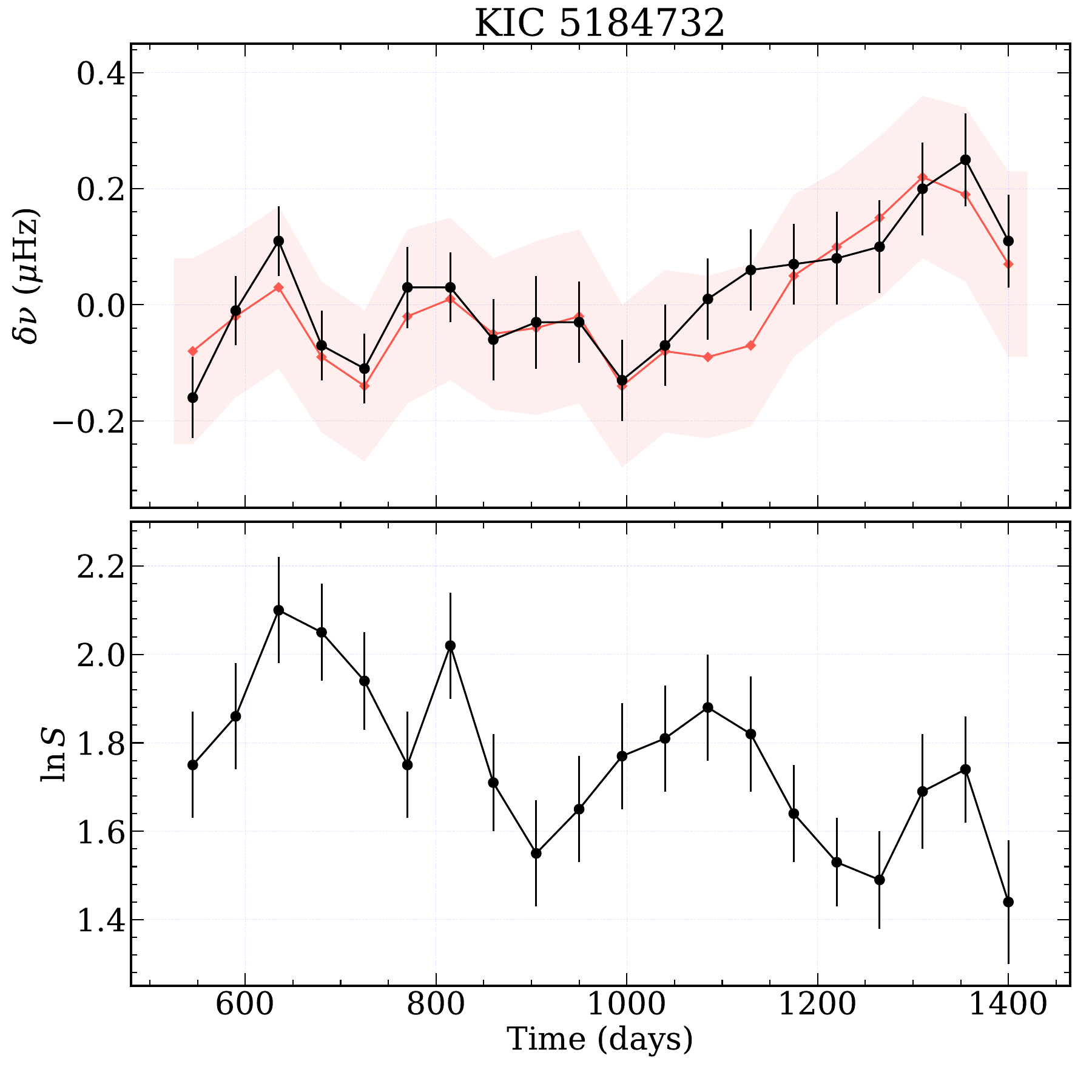}
 \caption{Asteroseismic detection of magnetic activity in the two solar-like \textit{Kepler} targets KIC~8006161 and KIC~5184732. \textit{Top panel for each star:} Mean frequency shifts $\delta\nu$ averaged over all available p-modes as a function of time as measured by Bayesian peak-bagging (black circles) and as measured with the cross-correlation technique (red diamonds). \textit{Bottom panel for each star:} Logarithmic mode height $\ln S$ of the p-modes obtained from peak-bagging. This figure was produced based on the original data from \cite{Santos2018}.}
 \label{fig:santos2}
\end{figure}

The search for seismically detected magnetic activity (cycles) was then further expanded by \cite{Santos2018} to 87 solar-like stars, including the 66 stars from the \textit{Kepler} LEGACY sample \citep{Lund2017, Aguirre2017} and 25 solar-like KOI targets \citep{Campante2016}. They used a Bayesian peak-bagging technique on 90-day segments of the photometric time series to measure the shifts of individual p-mode peaks in the periodogram enabling them to analyse different harmonic degrees and even azimuthal orders for variations in their parameters. The results of two of the stars in their sample are shown in Fig.~\ref{fig:santos2}. The two top panels show the frequency shifts of KIC~8006161 in the top panel and the logarithm of the mode amplitudes obtained from Bayesian peak-bagging in the bottom panel. The results of KIC~5184732, a more typical example from the analysed sample regarding the magnitude of the uncertainties and amplitude of the variations, are shown in the two bottom panels. In both cases, there are significant and systematic variations in p-mode frequencies as well as in mode amplitudes. What is more, these variations are anti-correlated (as measured by Pearson's correlation coefficient $r$) with one another, at a level of $r=-0.791$ for KIC~8006161 and $r=-0.482$ for KIC~5184732, signifying magnetic activity is likely causing these variations. 

Indeed, KIC~8006161 is probably one of the most intensively studied stars from the \textit{Kepler} seismic sample due to its very significant asteroseismic signature of magnetic activity. This star is very similar to the Sun in mass and radius but has a metallicity which is about twice the solar value. \cite{Karoff2018} analysed spectroscopic observations from the MWO program of KIC~8006161 spanning almost 20 years. This uncovered an activity cycle with a period of approximately \unit[7.4]{yr}. Through its cycle, KIC~8006161 has a significantly higher variability in its photospheric activity proxy $S_{\mathrm{ph}}$ and its $S$ index than the Sun. \cite{Karoff2018} postulated the star's higher metallicity brings about a deeper convective envelope compared to the Sun. This, in turn, then causes stronger levels of activity. The authors also show that the \textit{Kepler} era of observations is coincident with the rising period of magnetic activity for KIC~8006161. This lends further support to activity being the root cause of the measured p-mode parameter variations.
Due to the exquisite quality of the \textit{Kepler} data and the very good signal-to-noise ratio of the p-mode peaks in the periodogram of KIC~8006161, \cite{2019MNRAS.485.3857T} were able to constrain its latitudinal distribution of active regions by remapping the observed frequency shift to the stellar surface. Their technique utilises that modes of different harmonic degrees differ in their sensitivity to the latitudinal distribution of the perturbation causing the frequencies to shift \citep{Moreno-Insertis2000, Chaplin2007}. The authors determined that KIC~8006161's active regions are distributed over a wider band of latitudes and are located at higher latitudes than for the Sun. Based on a model of the rotation profile and the rotational modulation of this star's \textit{Kepler} time series, \cite{2018A&A...619L...9B} constructed a butterfly diagram. In their result, KIC~8006161 exhibits spots at both low latitudes close to the equator and, during some periods, at higher latitudes around 40$^{\circ}$. 

Following the advice of \cite{Karoff2009}, \cite{Karoff2013} observed 20 Sun-like stars in the \textit{Kepler} field-of-view with the Nordic Optical Telescope (NOT) and determined their excess flux (surface flux arising from magnetic sources) and \textit{S} index. From the stars' \textit{Kepler} light curves, they measured the rotation periods and the small frequency separation, which they used to guide the target selection for their program. The stellar fundamental parameters were obtained using an asteroseismic modelling code. \cite{Karoff2013} found that the ten stars from their sample which have independent measurements of asteroseismic ages, rotation periods and excess flux follow the Skumanich relations \citep{Skumanich1972} reasonably well. Further and interestingly, they obtained a much stronger relation between asteroseismically determined stellar properties and the stars' excess flux than with their \textit{S} index. \cite{Karoff2019} subsequently analysed the full four years of NOT spectroscopic data (covering the complete \textit{Kepler} main mission 2009--2013) as well as the photometric variability and p-mode frequency shifts of these 20 stars. They detected a strong correlation between the different activity proxies only for a few targets, most notably for KIC 8006161. The authors attribute this to the rather sparse sampling of spectroscopic data and the relative shortness of the photometric time series compared to the expected length of activity cycles. While \cite{Karoff2013, Karoff2019} did not specifically look for or find new asteroseismic detections of stellar activity cycles, they showed how asteroseismology -- in conjunction with ground-based spectroscopic data -- can usefully inform research on stellar activity, activity cycles, and the investigation of age–rotation–activity relations of solar-like oscillators.

\begin{figure}
 \centering
 \includegraphics[width=0.6\textwidth]{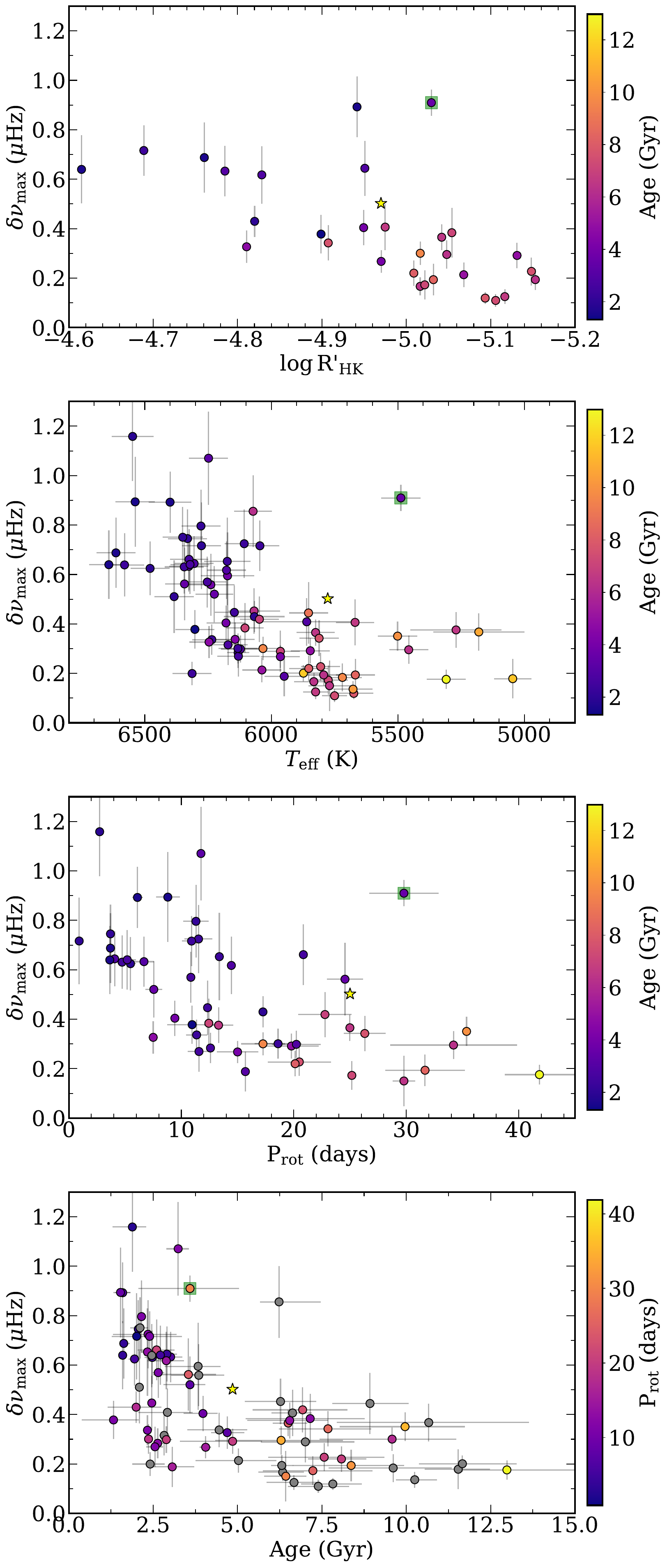}
 \caption{Frequency shift amplitudes $\delta\nu_{\mathrm{max}}$ of 75 \textit{Kepler} stars as a function of different stellar parameters: chromospheric activity level $\log R'_{\mathrm{HK}}$ (\textit{first panel}), effective Temperature $T_{\mathrm{eff}}$ (\textit{second panel}), rotation period $P_{\mathrm{rot}}$ (\textit{third panel}), and age (\textit{fourth panel}). The colours of the data points indicate the stars' ages, except for the bottom panel, where they encode rotation period. KIC~8006161 and the Sun are highlighted by the light green square and yellow star, respectively. This figure was produced based on the original data from \cite{Santos2019}.}
 \label{fig:santos1}
\end{figure}

\subsection{p-mode frequency variations and their relations to fundamental stellar parameters}\label{sec4.3}
\cite{Santos2019} investigated whether the amplitudes of the mean frequency shifts $\delta\nu_{\mathrm{max}}$ given by \cite{Santos2018} depend on fundamental stellar parameters. Parts of their results are reproduced in Fig.~\ref{fig:santos1}. They found that there is a strong correlation between the amplitude of the frequency shifts and the chromospheric activity level as measured by the $\log R'_{\mathrm{HK}}$ index with a Spearman rank correlation coefficient of $\rho=0.69$ (first panel of Fig.~\ref{fig:santos1}), supporting magnetic activity as the root cause of the measured shifts. As a function of effective temperature $T_{\mathrm{eff}}$, $\delta\nu_{\mathrm{max}}$ increases. This is not necessarily caused by an increase in the strength of magnetic activity with $T_{\mathrm{eff}}$ but is most likely caused by the increased mode sensitivity. Both, \cite{Metcalfe2007} (based on work done by \citealt{Dziembowski2004}) and \cite{Kiefer2019} used a simple scaling model for the frequency shifts, assuming that they are proportional to the depth of the perturbation causing them (magnetic activity), the p-modes' inertia, and the strength of the activity. The relations which they deducted reproduce the course of $\delta\nu_{\mathrm{max}}$ seen in the second panel of Fig.~\ref{fig:santos1} reasonably well. As can be seen in the bottom two panels of Fig.~\ref{fig:santos1}, $\delta\nu_{\mathrm{max}}$ decreases as rotation period $P_{\mathrm{rot}}$ increases and as stars age. This is expected, as stellar activity as well as stellar rotation is known to decay as stars get older \citep[see, e.g,][]{Skumanich1972, 1981ApJ...250..276V, Noyes1984}.

As the frequencies of p-modes of unequal harmonic degrees are variably susceptible to perturbations, also the frequency separation ratios are subject to magnetic activity-induced variations. \cite{Thomas2021} investigated the bias caused by activity-perturbed separation ratios on the estimates for fundamental stellar parameters through stellar modelling pipelines. They determined that, for solar-like stars with activity levels similar to the Sun, the bias is typically less than \unit[0.5]{\%} for mass, but can affect estimates of stellar age by up to \unit[5]{\%} and core hydrogen content by up to \unit[3]{\%}. Stronger than solar activity levels consequently increases these errors, as will extreme inclination angles, because the separation ratios are more strongly perturbed in such a scenario. Similarly, \cite{PerezHernandez2019} found that the activity-induced variation of the small frequency separation, i.e., the frequency separation of consecutive quadrupole and radial modes, can cause misdetermination of stellar age by up to \unit[10]{\%} and of mass and radius by a few percent. However, they also find this variation can often be masked by filtering out surface effects from the mode frequencies.

As all p-mode frequencies shift with activity, the frequency of maximum oscillation amplitude $\nu_{\mathrm{max}}$ follows suit. In their study of the temporal variation of the solar $\nu_{\mathrm{max}, \odot}$, \cite{Howe2020} indeed found that it is positively correlated with the level of solar magnetic activity and that it changes by as much as $\simeq$\unit[25]{\textmu Hz} between solar activity minimum and maximum. As $\nu_{\mathrm{max}, \odot}$ is used in asteroseismic scaling relations, this shift can incur an error of up to \unit[0.8]{\%} and \unit[2.4]{\%} in the estimates of stellar radius and mass, respectively. 

\subsection{Theoretical groundwork and recent detection of interior magnetic fields}\label{sec4.4}
Before the advent of asteroseismology, there have been a number of theoretical developments in helioseismology dealing with the effects of internal magnetic field on the oscillation frequencies of the Sun. \cite{Gough1990} used perturbation theory for the calculation of the frequency shifts of p-mode multiplets caused by buried magnetic field distributions. Their framework was later expanded by \cite{Antia2000} and \cite{Baldner2009} for the investigation of long time series of solar p-mode splitting coefficients to yield information on magnetic field concentrations in the solar convective envelope. Utilizing another flavour of perturbation theory, \cite{Lavely1992} applied a quasi-degenerate perturbational ansatz to deduct a theoretical framework for the calculation of the effect of convection and elastic-gravitational asphericities on p-modes. This was later expanded to include the effect of sub-surface magnetic field concentrations by \cite{Kiefer2017b} and \cite{Kiefer2018}. Similarly, \cite{Hanasoge2017} and \cite{Das2020} deduct expressions describing the effect of Lorentz stresses on the coupling of solar oscillations.

These theories can also be applied to other solar-like stars -- albeit in a more limited fashion due to the spatially unresolved nature of stellar observations -- and as long as the physical boundary conditions with which the theory has been developed are respected. Using stellar structural and oscillation models, these theories can emulate the fingerprint of cyclic variations of magnetic field strengths and configurations in the stellar oscillation frequencies.

Over the last few years, several groups have pushed forward the theoretical description of and, consequently, the search for signatures of the interplay of internal magnetic fields and stellar oscillation. These recent studies largely focused on the impact of internal magnetic field on gravity or mixed-mode frequencies: \cite{2020A&A...638A.149V} calculated frequency shifts in gravity-mode pulsators at the end of the main-sequence and find that axisymmetric poloidal-toroidal fields stronger than \unit[$10^6$]{G} should be detectable from \textit{Kepler} time series, if these fields exist. \cite{Prat2020} predicted that oblique dipolar magnetic fields leave detectable signatures in the gravity mode periods by applying their theory to a magnetic, rapidly rotating and slowly pulsating B-type star.
Several recent studies investigated the effects of magnetic field distributions of various configurations on mixed modes in red giants \citep{2020MNRAS.496..620G, 2021A&A...647A.122M, 2021A&A...650A..53B, 2022A&A...667A..68B, Loi2020, 2021MNRAS.504.3711L} and $\gamma$ Dor and SPB stars \citep{Dhouib2022}.

In the wake of this flurry of recent theoretical studies, the existing data sets were reanalysed, looking for the predicted signatures of magnetic fields in the oscillation frequencies. Using \textit{Kepler} data, \cite{2022Natur.610...43L} detected strong magnetic fields in the cores of giant stars. For three hydrogen-shell burning giants, they measured asymmetries in the (mixed-mode) oscillation multiplets, which translate into magnetic field strengths of \unit[102$\pm$12]{kG} for KIC 8684542, \unit[98$\pm$24]{kG} for KIC 7518143, and an upper limit of \unit[41]{kG} for KIC 11515377. Recently, building on the technique presented by \cite{2022Natur.610...43L}, \cite{2023arXiv230101308D} seismically detected strong magnetic fields in the cores of 11 red giants, again using \textit{Kepler} data. These observational studies are thus far restricted to static magnetic fields, i.e., one data point in time without the detection of cyclic activity. However, they clearly show that -- given long enough data sets -- variable or cyclic magnetic field concentrations can be detected in the stellar interiors using asteroseismology.

\section{Activity cycles in photometric time series}\label{sec5}
Stellar activity cycles can also reveal themselves as periodic variations of stellar brightness (see, e.g., \citealp{1985ARA&A..23..379B}), just as the Sun's total irradiance varies slightly over the solar 11-year cycle (see, e.g., \citealp{2014A&A...570A..85Y}).  Using long-term observations of the $V$-band magnitude, \cite{2000A&A...356..643O} were able to detect starspot cycles for nine out of the ten stars in their sample of rapidly-rotating active stars (a mix of K and G stars, including main-sequence stars, subgiants, and giants) with data lengths between \unit[11--30]{yr}. They found that the detected photometric cycle lengths agree with those from other activity proxies found by other authors. Later, \cite{2002AN....323..361O} expanded this study to a baseline of up to \unit[34]{yr} and could confirm that cycle period depends on rotation rate.

Evaluating four-year-long photometric light curves from the \textit{Kepler} satellite, \cite{2014MNRAS.441.2744V} presented evidence for activity cycles for nine of the 39 fast-rotating late-type active stars they investigated. The cycles they detected have periods between \unit[300--900]{d}. \cite{2014MNRAS.441.2744V} used the temporal variation of the stars' rotation period as an indicator for the cycles. Using multi-decadal ground-based photometric data, this approach enabled \cite{2009A&A...501..703O} to detect activity cycles in at least 15 of the 20 active stars in their sample. \cite{2015A&A...583A.134F} found evidence for stellar cycles in the photometric data of 16 CoRoT FGK main sequence stars. These cycles follow the earlier-found relations between the length of the activity cycles and the stars' rotation periods. Further, in addition to the active and inactive branches in the $P_{\rm rot}$-$P_{\rm cyc}$ diagram proposed by \cite{BohmVitense2007}, they detected hints for a possible third branch for short cycles.

With between 16 and 27 years of Johnson $B$- and $V$-band photometry from the Automatic Photoelectric Telescope (APT) at the Fairborn Observatory in Arizona, \cite{2016A&A...588A..38L} investigated differential photometry from 21 young solar-type stars. They detected photometric activity cycles in nearly all of the targeted stars. Populating the $\log \frac{P_{\rm rot}}{P_{\rm cyc}}$-$\log \rm{Ro}^{-1}$ diagram as defined in \cite{Saar1999}, they could confirm the active and transitional activity branches and found that the transitional branch merges with the active branch at $\approx \log \rm{Ro}^{-1} = 1.42$, similar to what was reported by \cite{BoroSaikia2018A&A...620L..11B}.   Using time-frequency analysis, \cite{2019MNRAS.483.2748S} investigated the temporal variations in chromospheric ($S$ index) and photometric (differential photometry in $b$- and $y$-bands) of decades-long time series. They found that activity of the young rapidly rotating solar analogue HD 30495 (also see \citealp{2015ApJ...812...12E} for a detailed analysis of this star), as measured with these two time series are strongly correlated. They detected activity cycles in the 'mid-term' regime with a length of \unit[1.6--1.8]{yr} and confirmed a longer cycle with a period of $\approx$\unit[11]{yr}.

\cite{2014JSWSC...4A..15M} defined two simple measures of photometric activity levels: $S_{\rm ph}$ is the standard deviation of the complete time series and thus reflects an average level of activity. With $\langle S_{\rm ph, k}\rangle$, the standard deviation of the time series is calculated over $k\times P_{\rm rot}$. They identified $k=5$ to be a good value, which smoothes out variations by rotation sufficiently, while still leaving longer-cycle variations intact. These measures are both often-used in the investigation of space-photometry for activity \citep[e.g.,][]{2016A&A...596A..31S, Karoff2018, Mathur2019}.  \cite{2020AN....341..508D} show that stellar activity cycles can also be detected in Gaia photometric time series. For two Gaia targets, they present evidence of cyclic photometric variations with cycle lengths of $P_{\rm{cyc}}\approx\unit[500]{d}$ for Gaia~DR2~2925085041699059712 and of $P_{\rm{cyc}}=\unit[3262\pm125]{d}$ for Gaia~DR2~3246069594362282752.

The as-of-yet largest number of detections of photometric activity cycles was achieved by \cite{2017A&A...603A..52R}. They analysed the long-cadence \textit{Kepler} times series of 23601 stars. As a signature for photometric variability, and hence for varying levels of magnetic activity, they measured the variability amplitude within each \textit{Kepler} quarter ($\approx\unit[90]{d}$) as the difference between the 5th and 95th percentiles of the light curve. They found amplitude periodicities in 3203 stars with cycle periods between $\unit[0.5]{yr}<P_{\rm{cyc}}<\unit[6]{yr}$ with stellar rotation periods between $\unit[1]{d}<P_{\rm{rot}}<\unit[40]{d}$. Interestingly, they confirmed, by folding all of the detected cycles, that the average shape of the stellar activity cycle deviates from a perfect sine, in particular during epochs of maximal and minimal activity. No dependence on stellar effective temperature was detected for this behaviour. The detections are scattered around the inactive (I) branch in the $P_{\rm{cyc}}$-$P_{\rm{rot}}$ diagram (cf., \citealp{Saar1999, BohmVitense2007}) with only few detections on the active (A) and short-cycle (S) branches. The authors propose that this may be due to the strong sensitivity of \textit{Kepler} photometry to spots and plages in the photosphere, while other studies, which have detected the A and S branches, used chromospheric activity indicators.

The \textit{Kepler} satellite recorded 53 full-frame images (FFIs) over the course of its main mission. \cite{2017ApJ...851..116M} used these FFIs to investigate a set of 3845 stars (F7 to G4) for signs of long-term photometric variability. For approximately $\unit[10]{\%}$ of their targets, 463 stars, \cite{2017ApJ...851..116M} observed significant ($>3\sigma$) brightness variations over the \textit{Kepler} mission. By eye, they detected apparently complete cycles for 28 stars. Further, they identified the range of rotation periods during which the transition from spot- to facula-dominated variability occurs, to lie between $\unit[15]{d}<P_{\rm{rot}}<\unit[25]{d}$. Also, the detected cycles appear to follow the A and I branches in the $P_{\rm{cyc}}$-$P_{\rm{rot}}$ diagram.

\cite{2020ApJ...901...14B} generated a large number of light curves based on starspot models in order to understand degeneracies in these light curves affected by starspots with varying lifetimes and distributions as well as underlying global and differential rotation. In light of their study, caution must be taken when interpreting short-term cyclic behavior in photometric light curves as the signature of possible activity cycles: such cycle-like behaviour can be the result of random fluctuations. The authors urge to reconsider past identifications of short-term activity cycles, such as those reported by, e.g., \cite{2014MNRAS.441.2744V, 2017A&A...603A..52R}, and encourage to scrutinize detections of cycles, which are solely based on variations in photometric time series, more carefully in future analyses.

\section{Summary and future prospects}\label{sec6}

In this review, we have presented the latest results on understanding stellar magnetic cycles on stars other than the Sun using the same diagnostics as are commonly used to quantify the solar cycle. Over the 11-year solar cycle, changes in the internal structure of the Sun are observed in acoustic oscillations (p-modes), while on the Sun's surface or the photosphere, the 11-year evolution of activity is evident from the patterns of spot emergence with ever decreasing latitudes as the cycle progresses. At solar activity maximum there is the largest number of starspots, and the Sun's large-scale magnetic field geometry is complex. In the Sun's outer atmospheric layers, for example in the chromosphere and corona, variations in its \textit{S} index and X-rays are co-incident with activity maximum. 

While the spatial resolution and time cadence of stellar observations is several orders of magnitude lower than observations of the Sun, we can use the same multi-wavelength diagnostics to quantify stellar magnetic activity. These are namely Doppler and Zeeman-Doppler imaging to map photospheric features such as dark, bright and magnetic spots, the \textit{S} index and X-ray observations as diagnostics of the stellar chromosphere and corona. In particular, we now have the capability to observe stars over time bases that are comparable to the length of the solar cycle. 

Observations of solar-like cycles have been observed on other stars, where the large-scale magnetic field geometry, including polarity reversals, is co-incident with \textit{S} index activity minimum and X-ray variations. For stars that have the same mass but are very much younger than the Sun, having just arrived on the zero-age main-sequence, they show more extreme levels of activity without any cyclic behaviour. As the stars age, the variations in magnetic activity start to become more cyclic like what we observe on the Sun, and could even have superimposed cycles. Long timespans of S-index monitoring show that cycles on solar-mass stars older than the Sun start to become lower in amplitude before eventually disappearing as the stars evolve off the main-sequence.  This is consistent with the recent results of \cite{Brown2022MNRAS.514.4300B} where 
chromospheric activity and variability is shown to decrease together with the toroidal field strength as stars evolve through their main-sequence lifetimes.  This is also in agreement with the results of \cite{Radick2018ApJ...855...75R} where they report that young stars have an inverse correlation between photometric brightness and Ca II emission, while more evolved main-sequence stars tend to show a direct correlation.  
Looking to the future, more long-term monitoring of the large-scale magnetic field geometry and S-index of stars that are both younger and older than the Sun, but with a similar mass, will provide us with an important insight into exactly when activity patterns start to become more regular and when they disappear. Similarly, the systematic observations of young, middle-aged and old stars with masses lower than the Sun will also provide us with an insight into how crucial stellar mass is for the stars' internal dynamo generation mechanisms.

Asteroseismology of stellar activity and activity cycles will take the next step forward with ESA's exoplanet-finding PLATO mission \citep{2014ExA....38..249R}. PLATO will observe two Long-duration Observation Phase (LOP) fields for two
years each. The exact position of these fields will be fixed two years before the targeted 2026 launch \citep{2022A&A...658A..31N}. At least 15000 dwarf and subgiant stars in the spectral type range F5--K7 with magnitudes $V\leq 11$ will be observed in the stellar sample with the highest priority at a time cadence of \unit[25]{s} (sample P1, see \citealp{2017EPJWC.16001003G, 2021A&A...653A..98M}). Another at least 1000 stars (dwarfs and subgiants, F5--K7) with $V\leq 8.5$ will be observed during a LOP (sample P2). This sample's higher brightness increases the feasibility of ground-based follow-up observations. Additional $\geq$245000 stars (sample P5, dwarfs and subgiants, F5--K7) with $V\leq 13$ will be observed with a lower signal-to-noise ratio than those in P1.
The duration of two consecutive years is not optimal for seismic studies of complete activity cycles, as shown in Sect.~\ref{sec4}. PLATO will still most likely expand the number of seismic targets with detected signatures of magnetic activity considerably and thus help improve our understanding of stellar dynamos, cycles, and how these depend on fundamental stellar parameters. If the PLATO mission should be extended and the same LOP be revisited, this would substantially increase the potential of the seismic detection and probing of activity cycles.

\bibliography{cycles.bib}


\backmatter

\bmhead{Supplementary information}


\bmhead{Acknowledgments}

We thank Stephen Marsden for assistance in reformatting Figure 2.  SVJ acknowledges the support of the German Science Foundation (DFG) priority program SPP 1992 `Exploring the Diversity of Extrasolar Planets' (JE 701/5-1).  The research leading to these results has received funding from the European Community's Seventh Framework Programme (FP7/2013-2016) under grant agreement number 312430 (OPTICON).

\section*{Declarations}

There are no conflicting interests to declare.



\end{document}